\documentclass[aps,prb,showpacs,reprint, superscriptaddress]{revtex4-1}

\usepackage{graphicx}
\usepackage{color}
\usepackage{amsmath}
\usepackage{bbm}

\input epsf

\begin{document}

\title{Fulde-Ferrell state induced purely by the orbital effect in a superconducting nanowire}
\author{Pawe{\l} W\'ojcik}
\email{pawel.wojcik@agh.edu.pl}
\affiliation{AGH University of Science and Technology, Faculty of
Physics and Applied Computer Science, 30-059 Krakow, Poland
Al. Mickiewicza 30, 30-059 Krakow, Poland}
\author{Micha{\l} Zegrodnik}
\email{michal.zegrodnik@agh.edu.pl}
\affiliation{Academic Centre for Materials and Nanotechnology, AGH University of Science and Technology, Al. Mickiewicza 30, 30-059 Krakow,
Poland}
\author{J\'ozef Spa\l ek}
\email{ufspalek@if.uj.edu.pl}
\affiliation{Marian Smoluchowski Institute of Physics, 
Jagiellonian University, ul. \L ojasiewicza 11,
30-348 Krak\'ow, Poland}
\affiliation{Academic Centre for Materials and Nanotechnology, AGH University of Science and Technology, Al. Mickiewicza 30, 30-059 Krakow,
Poland}

\date{10.05.2015}

\begin{abstract}
We show that the Fulde-Ferrell (FF) phase may appear as a sole result of the
orbital effect in a cylindrical metallic nanowire. Namely, in the external
magnetic field the two-fold degeneracy with respect to the orbital magnetic quantum number $m$ is lifted, what leads to a Fermi wave vector mismatch
between the subbands with opposite orbital momenta in the paired state. This
mismatch can be compensated by the nonzero total momentum of the Cooper
pairs created by electrons from the split subbands what results in the
formation of the FF phase. In this manner, a transformation of the orbital motion into a linear supercurrent parallel to the applied field is taking place. With the increasing magnetic field
a series of FF stability regions appear, in between which the standard
BCS superconducting phase is stable. For the sake of completness, we show, that the inclusion of the Zeeman term in the model does not change the picture qualitatively, particularly if larger $m$ states contribute essentially to the Fermi-surface splitting. A brief but important note concerning the possibility of steering the supercurrent by an applied magnetic field parallel to it, is also provided.
\end{abstract}


\maketitle
\section{Introduction}
According to the original concept by Fulde and Ferrell\cite{Fulde1964} (FF),
as well as by Larkin and Ovchinnikow\cite{Larkin1964} (LO), a superconducting phase with nonzero
center-of-mass momentum of the Cooper pairs can be induced by the Zeeman spin splitting of the
Fermi surface which appears in the external magnetic field. The Fermi wave vector mismatch, caused
by this Fermi surface splitting, is detrimental to the pairing, but can be compensated by the emerging nonzero
total-momentum of the Cooper pairs. This allows for the superconducting phase to persist in magnetic
fields substantially higher than the second critical field $H_{c2}$. In order to make it possible for the resultant non-zero
momentum Cooper pairing to appear, several requirements are to be met.
First of all, the Maki parameter\cite{SaintJames1969} should be large
meaning that the Pauli paramagnetic effect has to be strong relative to
the orbital pair-breaking mechanism.\cite{Gruenberg1966} Moreover, the system has to be very clean
as the FFLO phase is easily destroyed by the presence of impurities.\cite{Takada1970} Because of
those rather stringent conditions there are not many examples for such nonzero-momentum pairing to
occur. Until now, the experimental signs of the FFLO state have been suggested for the organic
superconductors\cite{Singleton2000,Tanatar2002,Uji2006,Shinagawa2007} and the heavy fermion compound
CeCoIn$_5$.\cite{Bianchi2003,Kumagai2006,Correa2007} The quasi-two-dimensional structure of those
systems leads to a strong reduction of the orbital pair breaking in the applied in-plane magnetic
field, as well as to nesting properties of the quasi-2D Fermi surface. Both of
these features are expected to stabilize the FFLO phase. 

An indirect experimental evidence of nonzero-momentum pairing has been recently reported for a
system consisting of ultracold $^6$Li atoms trapped in an array of one dimensional
tubes.\cite{Liao2010,Sun2012,Sun2013} One should also note the ongoing theoretical investigations
regarding the appearance of the FFLO state in other systems such as heavy fermion superconductors
with spin-dependent masses,\cite{Kaczmarczyk2010,Maska2010}
iron pnictides,\cite{Ptok2013_1,Ptok2014_1} 
ultracold fermionic atoms,\cite{Wu2013,Liu2013,Dong2013} as well as layered\cite{Croitoru2012,
Croitoru2012b, Croitoru2013} and quasi-one dimensional\cite{Croitoru2014} superconductors.

We consider here the case of a metallic cylindrical nanowire and demonstrate that the orbital effect, which
so far has been regarded as detrimental to the FFLO phase formation, can in fact induce the nonzero-momentum paired state. This situation is facilitated in a unique manner by the transfer of kinetic
energy from the rotational to the translational degrees of freedom. Namely, the rotational degrees
of freedom lead to the Fermi surface splitting since in the applied field the states with the
magnetic quantum numbers $m$ and $-m$ have opposite magnetic moments what in turn results in
opposite energy contributions. The Fermi wave-vector mismatch created by the splitting is transferred into the nonzero total momentum of the Cooper pairs along the nanowire inducing the FF phase formation.
We also provide an argument for a feasibility of such state observability.

The structure of the paper is as follows. In the next Section we determine the quasiparticle states in the wire with nontrivial azimutal states (i.e., in the direction perpendicular to its length) within a modified Bardeen-Cooper-Schriffer (BCS) approach, in which the superconducting gap acquires an explicit angular momentum dependence. We also specify explicitly how the Fermi wave vector mismatch arises solely from the angular-momentum induced Fermi surface splitting. The self-consistent equation for the gap, together with that determining the chemical potential for the fixed carrier concentration, are also specified. In Section III we discuss our results, as well as predict concrete experimentally verifiable effects of sandwiching the FF states in the sequence of the BCS-type states with the increasing applied magnetic field. Finally, Sec. IV is 
devoted to a brief summary and a general discussion to what extent the proposed effect can be useful in creating/modifying the supercurrent in quantum-electronics devices. Additionally, some details concerning the derivation of the self-consistent equations for the superconducting gap and the chemical potential are deferred to the Appendix A, whereas the elementary determination of the superconducting parameters is provided in Appendix B, together with a direct interpretation of our results.

\section{Model}
We start with introducing the cylindrical coordinates $(r,\varphi,z)$ and choosing the gauge
for the vector potential as
$\mathbf{A}=(0,eH_{||}r / 2,0 )$, where the magnetic field $H_{||}$ is parallel to the nanowire
axis. Then the single-electron Hamiltonian in the cylindrical nanowire then has the following form
\begin{equation}
 \hat{H}_0=\frac{\hbar ^2}{2m_e} \left [ -\frac{1}{r}
\frac{\partial}{\partial r} r \frac{\partial}{\partial r} + \left ( -
\frac{i}{r} \frac{\partial}{\partial \varphi} + \frac{eH_{||}r}{2\hbar}
\right ) ^2 - \frac{\partial ^2}{\partial z ^2} \right ]\;,
\label{ham}
\end{equation}
where $m_e$ is the electron mass and $e$ is the electron charge. To underline the role of orbital
degrees of freedom we have neglected the Zeeman term which complicates uneccesarily the analysis
(the argument remains of the same type). For the sake of completeness, at the end of Sec. III we
show, that the inclusion of the Zeeman term does not change the results qualitatively. Hamiltonian 
(\ref{ham}) is simplified further due to the negligible role of the diamagnetic term $\sim \mathbf{A}^2$ which for
nanowires is one order of magnitude lower than that of the order parameter. In the nanowire geometry
the quantization of the single-electron energy into a series of subbands appears. These subbands are
indexed by the orbital magnetic $m$ and radial $j$ quantum numbers, what leads to the dispersion
relations
\begin{equation}
\xi _{k,m,j}=\frac{\hbar ^2 \gamma _{m,j} ^2}{2m_eR^2} + \frac{\hbar ^2
k^2}{2m_e} + m\mu_BH_{||} - \mu\;, 
\label{eq:single_el_ene}
\end{equation}
where $k$ is the particle momentum along the nanowire ($z$ axis),
$\gamma _{m,j}$ is the $j$-th zero of the $m$-th order Bessel function, $R$ is the nanowire radius, and
$\mu$ is the chemical potential. The corresponding single electron wave
functions have the form
\begin{equation}
\phi_{kmj}(r,\varphi,z)=\frac{1}{2\pi}\tilde{J}_{mj}(r)e^{im\varphi}e^{
ikz}\;,
\label{eq:wave_func}
\end{equation}
where  $\tilde{J}_{mj}(r)$ is defined as
\begin{equation}
\tilde{J}_{mj}(r) = \frac{\sqrt{2}}{R J_{m+1}(\gamma _{m,j})} J
_{m}\bigg(\frac{\gamma _{m,j}}{R}r\bigg)\;,
\end{equation}
with $J _{m}(r)$ being the $m$-th order Bessel function. One should note that
$\xi_{k,m,j}=\xi_{k,-m,j}$ for $H_{||}=0$.

\begin{figure}[h!]
\begin{center}
\includegraphics[scale=.45]{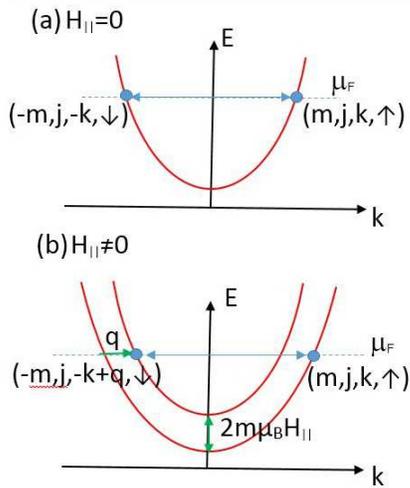} 
\end{center}
\caption{Schematic representation of Cooper paired states in the nanowire. (a) BCS case, (b)
FF pairing. The horizontal arrow connects the pair partners, for which the sets of quantum numbers are specified in the brackets.}
\label{fig1}
\end{figure}

Next, we include the effective electron-electron attraction term in the model and perform the BCS
approximation with the possibility of non-zero momentum pairing (with
the momentum $q$ along the $z$-axis). The BCS Hamiltonian is then of the form
\begin{eqnarray}
\hat{H}&=& \sum_{kmj}\xi_{k,m,j}\hat{n}_{k,m,j} + \sideset{}{'}\sum_{mjq}
\frac{|\Delta_{mjq}|^2}{V} \nonumber \\ 
&+& \sideset{}{'}\sum_{kmjq}\big(\Delta_{mjq}\hat{c}^{\dagger}_{k,m,
j\uparrow}\hat{c}^{\dagger}_{-k+q,-m,j\downarrow}+H.C.\big)\;,
\label{eq:Hamiltonian_BCS}
\end{eqnarray}
where $V$ is the coupling constant and
$\Delta_{mjq}$ is the superconducting (SC) gap defined by 
\begin{equation}
\Delta _{mjq}=\frac{V}{4 \pi^2} \sideset{}{'}\sum _{km'j'} C_{mjm'j'}
\langle\hat{c}_{-k+q,-m',j\downarrow}\hat{c}_{k,m',j\uparrow}\rangle
\label{eq:delta_def}\;,
\end{equation}
with the gap modulation
\begin{equation}
C_{mjm'j'} = \int_0^R dr \: \: r \tilde{J} _{mj}^2(r) \tilde{J}
_{m'j'}^2(r)\;.
\end{equation}
The primed summation means that $m$ runs over the nonnegative values only.
Additionally, for the summation over $k$, the pairing appears only in the range
$[\mu-\hbar\omega_D, \mu+\hbar\omega_D]$, where $\omega_D$ is the Debye frequency. From Eq.
(\ref{eq:Hamiltonian_BCS}) one can see that for $q=0$ the
pairing appears within each subband labelled with the radial quantum
number $j$ between particles with opposite spins, momenta, and orbital
momenta: ($k,m,j\uparrow$; $-k,-m,j\downarrow$). So the pairing contains both the translational and
the rotational degrees of freedom. In the applied field the degeneracy with respect to $m$ is
removed resulting in a shift between the subbands
corresponding to $m$ and $-m$  [cf. Eq. (\ref{eq:single_el_ene})]. In such situation, a
Fermi-wave-vector mismatch appears. However, this mismatch can be transferred into the nonzero
center-of-mass momentum of the Cooper pairs ($q\neq 0$) giving rise to
the FF phase induced purely by the orbital effect. This
process is represented in Fig.~\ref{fig1}.
For simplicity, we limit to the situation in which all the Cooper pairs
have a single momentum $q$ constituting the Fulde-Ferrell phase. \\
Diagonalization of (\ref{eq:Hamiltonian_BCS}) leads to the quasiparticle
energies
\begin{equation}
\begin{split}
  E^{\pm}_{kmjq}&=\pm\frac{1}{2}( \xi _{k,m,j}-\xi _{-k+q,-m,j})  \\
  &+ \sqrt{\frac{1}{4}( \xi _{k,m,j} + \xi _{-k+q,-m,j})^2 +\Delta
_{mjq} ^2}.
\end{split}
\label{eq:quasi_dis_rel}
\end{equation}
The set of self-consistent equations for the SC gap parameter
and the chemical potential have the corresponding form
\begin{eqnarray}
 \Delta _{mjq}&=&\frac{V}{4 \pi^2} \int dk \sideset{}{'} \sum _{mj}
C_{mj m'j'} \nonumber \\ 
&\times& \frac{\Delta _{mjq} \left [ 1- f(E^{+}_{kmjq}) -
f(E^{-}_{kmjq}) \right ]}{ \sqrt{( \xi _{m,j,k} + \xi _{-k+q,-m,j})^2 +4
\Delta_{mjq} ^2} }\;,
\label{eq:delta_self}
\end{eqnarray}
\begin{eqnarray}
 n_e&=&\frac{1}{\pi^2R^2} \int dk \sideset{}{'}\sum_{mj}  \int^R_{0}\:dr\:r
\big \{ |u_{kmjq}(r)|^2 f(E^+_{kmjq}) \nonumber \\  
&+& |v_{kmjq}(r)|^2[1-f(E^-_{kmjq})]\big \}\;,
\label{eq:mu_self}
\end{eqnarray}
where $n_e$ is the electron concentration and $f(E)$ is the Fermi-Dirac
distribution. One should note that Eqs. (\ref{eq:delta_def}) and
(\ref{eq:delta_self}) correspond to the SC gap in
reciprocal space.  Also, in the standard BCS theory the chemical potential is regarded as the same in the normal and superconducting phases. Here, as the confinement can significantly affect the chemical potential, we prefer to adjust it in each of the phases (normal and superconducting) separately, as it should be in principle. Details concerning the derivation of the self-consistent equations are contained in the Appendix at the end of the paper. 

The confinement in the radial directions of the
nanowire leads to the $r$ dependence of the SC gap which is determined by 
\begin{eqnarray}
 \Delta_q(r)&=&\frac{V}{4\pi ^2} \int
dk \sideset{}{'}\sum_{mj} v^{\star}_{kmjq}(r)u_{kmjq} (r) \nonumber \\
 &\times&\left [ 1- f(E^{+}_{kmjq}) - f(E^{-}_{kmjq}) \right ],
 \label{eq:delta_space}
\end{eqnarray}
where $u_{kmjq}(r)=U_{kmjq}\tilde{J}_{mj}(r)$ and
$v_{kmjq}(r)=V_{kmjq}\tilde{J}_{mj}(r)$, with $U_{kmjq}$, $V_{kmjq}$
being the Bogoliubov coherence factors, which in turn are elements
of the diagonalization matrix of (\ref{eq:Hamiltonian_BCS}).\\
The SC gaps $\Delta_{mjq}$ and the chemical potential are
obtained by solving Eqs. (\ref{eq:delta_self})
and (\ref{eq:mu_self}) numerically, whereas the wave-vector $q$ is determined
by minimizing the free energy of the system.\cite{Kosztin1998}
Additionally, by using Eq. (\ref{eq:delta_space}) the spatial dependence
of the gap can be determined explicitly.
\section{Results}
Calculations have been carried out for the following values of the
parameters: $\hbar \omega_D=32.31$~meV, $gN(0)$=0.18, where
$N(0)=\frac{mk_F}{2\pi^2 \hbar ^2}$ is the bulk density of states at
the Fermi level, $\Delta _{bulk}=0.25$~meV and $\mu _{bulk}=0.9$~eV
which corresponds to the electron density $n_e=3.88 \times 10^{18}$~cm$^{-3}$. The selected
parameter values correspond to Al superconductor.\cite{Shanenko2006} 
\begin{figure}[h!]
\begin{center}
\includegraphics[scale=.35]{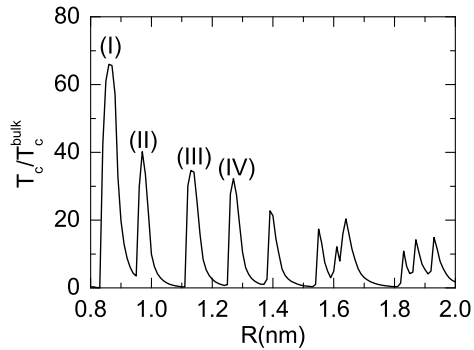} 
\end{center}
\caption{Superconducting critical temperature $T_c/T_c^{bulk}$ as a
function of nanowire radius $R$. The analysis of the unconventional
paired phase proposed in this work is carried out for $R$ corresponding
to the maxima labeled by I, II, III, and IV. Note that the same results have
been obtained in Ref. \onlinecite{Shanenko2006} (cf.
Fig. 1).}
\label{fig2}
\end{figure}
To determine the values of
$R$ which are appropriate for the analysis of the proposed
unconventional paired phase, we have calculated the nanowire radius
dependence of the critical temperature in zero magnetic field. The $T_c$
oscillations seen in Fig. \ref{fig2} have identical form as those
presented in Ref. \onlinecite{Shanenko2006} and result from the
electrons energy quantization due to the confinement effect. Maximum of
the $T_c$ appears each time an electron subband passes
through the energy window $[\mu_F-\hbar\omega_D, \mu_F+\hbar\omega_D]$. In what follows, the
possibility of non-zero momentum pairing due to the orbital effect is
analyzed for the values of the nanowire radius which correspond to the maxima (shape resonances)
labeled by I-IV in Fig.~\ref{fig2}.

\begin{figure}[!ht]
\begin{center}
\includegraphics[scale=.35]{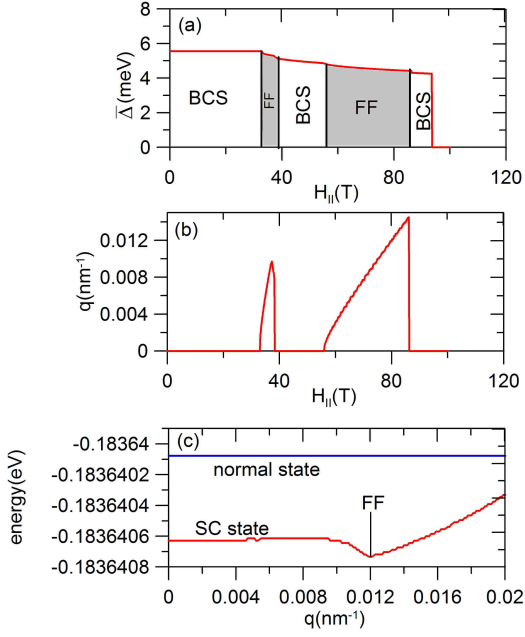} 
\end{center}
\caption{(a) Magnetic field dependence of the averaged superconducting
gap; (b) total Cooper pair momentum which minimizes the energy vs. $H_{||}$; (c) Cooper pair
momentum dependence of the energy in the paired and normal states for selected value of the field.
The energy minimum visible in (c) corresponds to the stability of FF phase. Results for
$R=0.97$~nm labeled by II in Fig.~\ref{fig2}.}
\label{fig3}
\end{figure}

In Fig. \ref{fig3} we show the magnetic field dependence of the averaged superconducting gap 
\begin{equation}
 \bar{\Delta}=\frac{1}{R} \int_{0}^R \:\Delta(r)\:dr,
\end{equation}
for the value of $R$ corresponding to $T_c$ maximum labeled by II. 
As one can see from Eq. (\ref{eq:single_el_ene}), the orbital pair breaking mechanism caused by the
Fermi surface splitting has
different magnitude for bands corresponding to different values of $|m|$. As a
result, the zero temperature superconducting-to-normal metal transition
driven by the magnetic field occurs as a cascade of jumps in the order
parameter as displayed in Fig.~\ref{fig3}(a). It has been
reported in Ref.~\onlinecite{Shanenko2008} that each jump corresponds to
Cooper pair breaking in a subsequent subband. Here, we show that the Fermi wave vector mismatch,
which appears in individual bands due to the orbital effect, can be compensated by nonzero
center-of-mass momentum of the Cooper pairs leading to the FF phase appearance. It is shown in Fig.
\ref{fig3}(a) that the stability of the FF phase appears for certain ranges of $H_{||}$ in between
which the BCS phase is stable. In Fig. \ref{fig3}(c) we show the minimum in the $q$-dependence of
the system energy for the selected value of external magnetic field which corresponds to the
stability of the FF state.

\begin{figure}[!ht]
\begin{center}
\includegraphics[scale=.35]{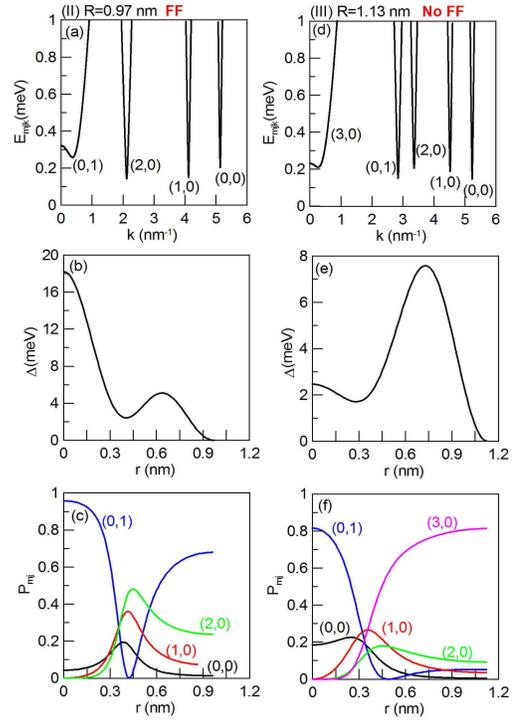} 
\end{center}
\caption{(a,d) Quasi-particle subbands labeled by $(m,j)$ participating in
SC state; (b,e) $r$-dependence of the
SC gap $\Delta(r)$; (c,e) contributions to the
paired phase coming from different subbands. Results are for
$R=0.97$~nm (II - left panels) and $R=1.13$~nm (III - right panels), both for $H_{II}=0$.}
\label{fig4}
\end{figure}

The structure of the FF and BCS stability ranges (cf. Fig.~\ref{fig3})
is caused by the multiband nature of the system. It should be
noted that the contributions to the paired phase coming from different
subbands, which is defined as (cf. Eq. (\ref{eq:delta_space}))
\begin{equation}
\begin{split}
 P_{m,j}(r)&=\frac{V}{4\pi\Delta_q(r)}\int
dk\: v^{\star}_{kmjq}(r)u_{kmjq} (r) \nonumber \\
 &\times\left [ 1- f(E^{+}_{kmjq}) - f(E^{-}_{kmjq}) \right ],
\end{split}
\end{equation}
may vary significantly. In Fig.~\ref{fig4}(a-c) we present the quasiparticle
branches for subbands ($m$,$j$) participating in the SC state,
as well as the
$r$-dependence of the SC gap for the selected radius $R$ in zero field.
By analyzing $\Delta(r)$ [Fig.~\ref{fig4}(b)] and $P_{m,j}(r)$ [Fig.~\ref{fig4}(c)] one can conclude that for the
selected resonant radius, there is a leading
quasiparticle branch e.g. $(0,1)$ for $R=0.97$nm (corresponding to the energy gap
$\Delta_{0,1}$) that controls the superconducting gap enhancement. The dominant contribution to the
SC state, which comes solely from single branch is a characteristic feature of each resonant
radius.\cite{Shanenko2006} The critical field, $H_{||c}$, at which the transition SC-normal metal
appears, is mainly determined by such dominant energy gap. The remaining quasiparticle branches are
responsible for the appearance of the FF-state stability ranges seen
in Fig. \ref{fig3}(a). Their influence on the formation of the FF phase can be explained as follows.
As already mentioned, the impact of the applied magnetic field on different subbands is dependent on
the orbital quantum number $|m|\neq 0$.
The larger $|m|$, the faster the Fermi wave vector mismatch rises with increasing $H_{||}$ (cf. Eq.
\ref{eq:single_el_ene}). In effect, the depairing in the subbands with different $|m|$ takes place
for different $H^m_{||}$. Close to $H^m_{||}$ the Fermi wave vector mismatch between the paired
electrons $|(k,m,j,\uparrow),(-k,-m,j,\downarrow)\rangle$ is compensated by the
non-zero total momentum of the Cooper pairs leading to the stable FF phase. However, upon further
increase of the field it is not possible to adjust the vector $q$ so as to sustain the pairing in
such subband and the Cooper pairs in this subband break up leading to a decrease of the energy gap
and standard BCS pairing reentrance. This allows to formulate the conditions for the appearance of
the FF stability regimes induced by the orbital effect in nanowires. Namely, the FF phase appears
only if the dominant quasi-particle branch has the orbital quantum number lower than the
other superconducting branches excluding those with $m=0$. \textit{The number of FF stability
regions is equal to the number of the subbands which fulfill this criterion.}
From Fig.~\ref{fig4} one can see that for $R=0.97$ nm there are two subbands $(1,0)$ and $(2,0)$
(subbands with $m=0$ are excluded) with a larger quantum number than the dominant subband $(0,1)$.
This results in two FF stability regions shown in Fig.~\ref{fig3}(a) with different slopes of
$q(H_{II})$, (Fig.~\ref{fig3}(b)) depending on the orbital quantum number $|m|$ corresponding to 
branches responsible for the $q\neq 0$ paired phase.
\begin{figure}[h!]
\centering
\epsfxsize=85mm 
{\epsfbox[21 272 591 585]{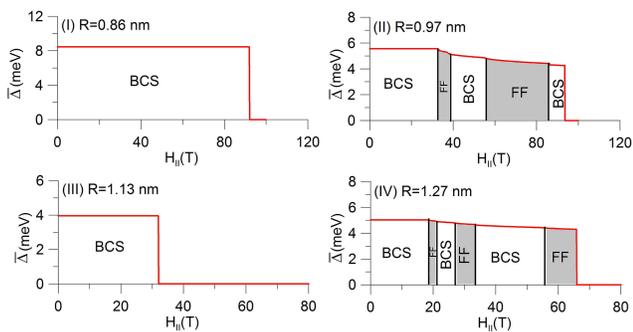}}
\caption{Magnetic field dependences of the averaged SC gap for four nanowire radii corresponding to
resonances I-IV, respectively.}
\label{fig5}
\end{figure}
As one can see in Fig. \ref{fig5}, the FF stability does not always
appear with increasing field. Such situation takes place when the dominant band corresponds to
large value of $|m|$ and the depairing in this band appears before it does in other quasiparticle branches.
The quasiparicle dispersion relations in such case ($R=1.13$~nm, (III)) are shown in Fig. \ref{fig4}
together with the corresponding $r$ dependence of the averaged gap parameter and the contributions
to the pairing coming from particular subbands. As one can see, for $R=1.13$ nm
the largest contribution to the pairing originates from the subband with
$m=3$ and $j=0$. Due to relatively large value of $m$, the corresponding
quasiparticle branch reaches zero energy before any other and as a
result, the FF state induced by the less important branches does not
appear. \\

\begin{figure}[h!]
\epsfxsize=85mm 
{\epsfbox[21 272 591 585]{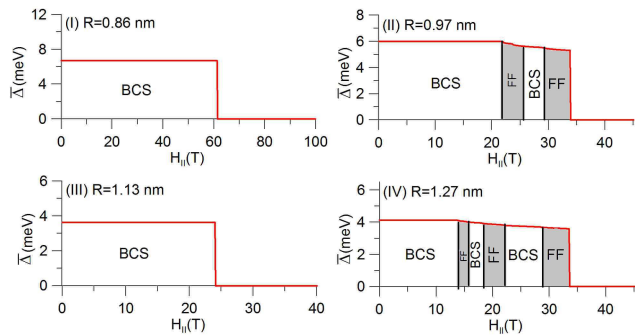}}
\caption{Magnetic field dependences of the averaged SC gap for four nanowire radii (I)-(IV)
calculated with the inclusion of the spin Zeeman effect.}
\label{fig6}
\end{figure}

\noindent \textbf{\textit{Zeeman effect}}.
To underline the role of the orbital effect in the creation of the FF state in nanowires, the
results presented so far did not include the influence of the spin Zeeman term. However, it should
be noted that the Pauli paramagnetism is the second factor being able to create the FF
phase in nanowires. The extension of our theoretical model for the case with the spin
Zeeman effect included, leads in a straightforward manner to the spin-dependent dispersion relations in the form analogous to Eq. (\ref{eq:single_el_ene}), namely
\begin{equation}
\xi _{k,m,j,\sigma}=\frac{\hbar ^2 \gamma _{m,j} ^2}{2m_eR^2} + \frac{\hbar ^2
k^2}{2m_e} + m\mu_BH_{||} + \sigma\mu_BH_{||}- \mu\;, 
\label{eq:single_el_ene_zem}
\end{equation}
where $\sigma=\pm 1$ for spin-up and down electrons, respectively.
Diagonalization of (\ref{eq:Hamiltonian_BCS}) with the dispersion relations
(\ref{eq:single_el_ene_zem}) leads to the quasiparticle energies in the form
\begin{equation}
\begin{split}
  E^{\pm}_{kmj}&=\pm\frac{1}{2}( \xi _{k,m,j,\uparrow}-\xi _{-k+q,-m,j,\downarrow})  \\
  &+ \sqrt{\frac{1}{4}( \xi _{k,m,j,\uparrow} + \xi _{-k+q,-m,j,\downarrow})^2 +\Delta
_{mjq} ^2},
\end{split}
\label{eq:dis_rel_zeeman}
\end{equation}
In Fig.~\ref{fig6} we present the stability diagrams calculated for the four nanowire radii
I-IV, with the inclusion of the spin Zeeman effect.
 As can be seen the obtained results do not differ qualitatively from the corresponding ones presented in Fig.
\ref{fig5}. The FF stability regions induced by the orbital effect are still clearly
visible. Nevertheless, as one can expect, the Zeeman effect reduces the value of the critical
magnetic field (cf. Figs. \ref{fig5} and \ref{fig6}). Its impact is crucial especially for resonances governed by the low orbital quantum numbers $m=0,1$; e.g. for $m=0$ the orbital effect
is not operative and only the Pauli paramagnetism determines the value of the corresponding critical magnetic field. By comparing Figs. \ref{fig5} and \ref{fig6} one can see that the largest reduction of the
critical field is obtained for $R=0.97$~nm which corresponds to the resonance governed by the state
with $m=0$ (cf. Fig.~\ref{fig4}). It should be noted that the inclusion of the spin Zeeman effect
does not alter the conclusion that for the states with higher orbital quantum number, which participate in
the formation of the superconducting phase (cf. Fig.~\ref{fig4}), the depairing takes place for the
lower values of the magnetic field giving raise to the stable FF phase. For the resonances governed
by the higher orbital quantum number $m$ the spin Zeeman effect does not play the crucial role and to
the first approximation it is negligible. For example, the reduction of the critical field for $R=1.13$~nm,
governed by $m=3$, is much less pronounced.

\section{Summary and outlook}
The orbital effect (i.e. the Fermi-surface dependence on the orbital magnetic quantum number, $m$) in cylindrical superconducting nanowires results in the Fermi wave-vector
mismatch which can be transferred to the nonzero linear momentum of the Cooper pairs. By solving the
system of Bogoliubov-de Gennes equations numerically, we have shown that this effect leads to stability
of the FF phase in certain ranges of the applied magnetic field
oriented parallel to the nanowire. The nonzero total momentum of the Cooper pairs is induced to
sustain the pairing in the corresponding subbands what in turn leads to structure of alternating FF and BCS stability ranges as shown in Figs. \ref{fig3} and \ref{fig4}. 
To expose the role of the orbital effect in the formation of the FF phase, we have first neglected the usual
spin-Zeeman term in our approach. However, for the sake of completeness, calculations with the
inclusion of the Pauli paramagnetism have also been carried out subsequently and show no significant qualitative
differences with respect to the case without the spin Zeeman term included (cf. Figs. \ref{fig5} and
\ref{fig6}). 

The proposed mechanism of the FF phase formation induced by the orbital effect appears only if few
states with different quantum numbers participate in the superconducting phase. At the same time, the thicker the nanowire is, the more states take part in the pairing. As a consequence, our analysis is restricted to relatively thin nanowires. Nevertheless, by lowering the electron concentration we also decrease the number of states taking part in the pairing. As shown in one of our previous papers, \cite{Wojcik2014} this fact leads to the increase of $T_c$ oscillations as a function of sample size for nanofilms (the same holds true for nanowires). This fact allows us to suggest that for a superconductor with low concentration of carriers, e.g., for SrTiO \cite{Schooley1964}, the presented effect may be observed also for much thicker nanowires.

By carrying out the specific heat measurements one could detect if the phase transitions
between the FF and BCS phases, shown in Figs. \ref{fig5} and \ref{fig6}, appear with the increasing magnetic field. Moreover, one should note, that
in the FF phase the Cooper pairs have nonzero total momentum which in our case leads to a supercurrent flow along the nanowire induced by a parallel to it magnetic field. This fact can also be used as a signal in the experimental detection of the proposed effect, as well as can have practical consequences in quantum-electronic devices, in the sense that the current can be created and stimulated by magnetic field applied parallel to its direction.

For simplicity, we have analyzed nanowires with a uniform cross section. One should note that the
effect of sample nonuniform cross section can make the experimental observation of the FF state obscured, since the energy gain coming from the nonzero momentum pairing is
rather small [cf. Fig. \ref{fig3}(c)]. On the other hand, a strong enhancement of the $T_c$
value with respect to that for a bulk sample should make the effect observable for uniform wires in
the clean limit. 

Finally, a methodological remark is in place here. The quantum wires considered in this paper have a nontrivial dynamics in the dimensions perpendicular to its length. Therefore, such systems cannot be regarded as strictly one-dimensional. This is the reason why we do not consider them from the electronic point of view as Tomanaga-Luttinger liquids. The question whether superconducting 1D liquids can produce the proposed effect, is still open.

\section{Acknowlegement}
This work was financed from the budget for Polish Science in the years 2013-2015, project number:
IP2012 048572. M. Z. and J. S. acknowledge the financial support from the Foundation for Polish
Science (FNP) within the project TEAM. J. S. acknowledges also the Grant MAESTRO from the National
Science Center (NCN), No. DEC-2012/04/A/ST3/00342.

\section*{Appendix A: Effective BCS Hamiltonian for the nanowire}
In this Appendix we present some details of the effective BCS Hamiltonian derivation for the superconducting nanowire in an external magnetic field. Moreover, we also show how to obtain the self-consistent equations for the gap parameter and the chemical potential.

We start from the general form of the BCS Hamiltonian
\begin{equation}
 \begin{split}
 \hat{\mathcal{H}}&=\sum _{\sigma} \int d^3 r
\hat{\Psi}^{\dagger} (\mathbf{r},\sigma) \hat{H}_0
\hat{\Psi}(\mathbf{r},\sigma) \\ 
&+ \int d^3 r \left [ \Delta
(\mathbf{r})\hat{\Psi}^{\dagger}(\mathbf{r},\uparrow)
\hat{\Psi}^{\dagger}(\mathbf{r},\downarrow) +H.C. \right ]\\
& +\int d^3r \frac{|\Delta(\mathbf{r})|^2}{V},
 \end{split}
 \label{eq:BCS_general}
\end{equation}
where $\hat{H}_0$ is the single-electron Hamiltonian, in our case defined by (\ref{ham}), and the gap parameter in real space is given by
\begin{equation}
 \Delta(\mathbf{r})=-V \left < \hat{\Psi} (\mathbf{r},\downarrow)
\hat{\Psi} (\mathbf{r},\uparrow)  \right >.
\label{eq:gap_def}
\end{equation}
For the case of cylindrical nanowire we chose the field operators in the form
\begin{equation}
\begin{split}
 \hat{\Psi}(\mathbf{r},\sigma)=\sum_{kmj}
\phi_{kmj}(\mathbf{r})\:\hat{c}_{kmj\sigma},\\
\hat{\Psi}^{\dagger}(\mathbf{r},\sigma)=\sum_{kmj}
\phi^*_{kmj}(\mathbf{r})  \:
\hat{c}^{\dagger}_{kmj \sigma},
\end{split}
\label{eq:field_op}
\end{equation}
where $\phi_{kmj}(\mathbf{r})$ are the eigenfunctions of $\hat{H}_0$ and are given by Eq. (\ref{eq:wave_func}). For the case of superconducting nanowire the Cooper pairs are created by electrons with opposite both spins and angular momenta. Additionally, we assume that the pairs can have nonzero total momentum, $q$, along the nanowire. In such situation, by substituting Eqs. (\ref{eq:gap_def}) and (\ref{eq:field_op}) into Eq. (\ref{eq:BCS_general}), one obtains the effective Hamiltonian in the form defined by Eq. (\ref{eq:Hamiltonian_BCS}). Next, by introducing the composite vector operators $\mathbf{\hat{f}}^{\dagger}_{kmjq}=(\hat{c}^{\dagger}_{k,m,j\uparrow}, \hat{c}_{-k+q,-m,j\downarrow})$ and assuming that all the Cooper pairs have a single momentum $q$ (the Fulde-Ferrell phase), we arrive at the following form of the effective Hamiltonian
\begin{equation}
\begin{split}
\hat{H}&= \sideset{}{'}\sum_{kmj}\mathbf{\hat{f}}^{\dagger}_{kmjq}\mathbf{H}_{kmjq}\mathbf{\hat{f}}_{kmjq}+\sideset{}{'}\sum_{kmj}\xi_{-k+q,-m,j} \\  &+\sideset{}{'}\sum_{mj}
\frac{|\Delta_{mjq}|^2}{V},  
\end{split}
\label{eq:Ham_matrix_1}
\end{equation}
where
\begin{equation}
\mathbf{H}_{kmjq}=\left(\begin{array}{cc}
\xi_{k,m,j}& \Delta_{mjq} \\
\Delta^*_{mjq} & -\xi_{-k+q,-m,j} \\
\end{array} \right),
\label{eq:matrix_H}
\end{equation}
and $\Delta_{mjq}$ is defined via Eq. (\ref{eq:delta_self}).
Hamiltonian (\ref{eq:Ham_matrix_1}) can be diagonalized by the use of the Bogoliubov transformation to new qusiparticle operators $\hat{\alpha}_{k,m,j,q}$ and $\hat{\beta}_{k,m,j,q}$, which has the following form
\begin{equation}
\left(\begin{array}{cl}
        \hat{c}_{k,m,j\uparrow}\\
        \hat{c}^{\dagger}_{-k+q,-m,j\downarrow}\\
       \end{array}\right)=\left(\begin{array}{cc}
U_{kmjq} & V_{kmjq}  \\
-V_{kmjq} & U_{kmjq} \\
\end{array}\right)\left(\begin{array}{cl}
        \hat{\alpha}_{k,m,j,q}\\
        \hat{\beta}^{\dagger}_{k,m,j,q}\\
       \end{array}\right),
\label{eq:Bogolobov_trans}
\end{equation}
where
\begin{equation}
 \begin{split}
  U_{kmjq}=\frac{1}{2}\bigg(1+\frac{\xi_{k,m,j}+\xi_{-k+q,-m,j}}{\sqrt{(\xi_{k,m,j}-\xi_{-k+q,-m,j})^2+4\Delta_{mjq}^2}}\bigg),\\
  V_{kmjq}=\frac{1}{2}\bigg(1-\frac{\xi_{k,m,j}+\xi_{-k+q,-m,j}}{\sqrt{(\xi_{k,m,j}-\xi_{-k+q,-m,j})^2+4\Delta_{mjq}^2}}\bigg),
 \end{split}
\end{equation}
are the Bogoliubov coherence factors. The columns of the unitary transformation matrix from Eq. (\ref{eq:Bogolobov_trans}) are the eigenvectors of the matrix form of our Hamiltonian (\ref{eq:matrix_H}). After the diagonalization transformation, the Hamiltonian has the following form
\begin{equation}
\begin{split}
 \hat{H}&=\sideset{}{'}\sum_{kmj}(E^+_{kmjq}\hat{\alpha}^{\dagger}_{k,m,j,q}\hat{\alpha}_{k,m,j,q} +  E^-_{kmjq}\hat{\beta}^{\dagger}_{k,m,j,q}\hat{\beta}_{k,m,j,q})\\
 &+\sideset{}{'}\sum_{kmj}(\xi_{-k+q,-m,j}-E^-_{kmjq})+\sideset{}{'}\sum_{mj}
\frac{|\Delta_{mjq}|^2}{V},
\end{split}
\label{ham_diagonal}
\end{equation}
where $E^{\pm}_{kmjq}$ are given by Eq. (\ref{eq:quasi_dis_rel})

The self-consistent equation for the superconducting gap (\ref{eq:delta_self}) can be derived by substituting for $\hat{c}_{-k+q,-m,j,\downarrow}$ and $\hat{c}_{k,m,j,\uparrow}$ from Eq. (\ref{eq:Bogolobov_trans}) into Eq. (\ref{eq:delta_def}). Similarly, the equation for the chemical potential (\ref{eq:mu_self}) is obtained by inserting the creation and anihilation operators from (\ref{eq:Bogolobov_trans}) into the equation for the electron concentration
\begin{equation}
 n_e=\frac{2}{\Omega}\int\:dk\sum_{mj}\langle\hat{n}_{k,m,j} \rangle,
\end{equation}
where $\Omega$ is the system volume. It should be noted that with the use of the self-consistent equation (\ref{eq:delta_def}) one obtains the superconducting gap in reciprocal space. After solving the self-consistent equations, the position dependence of the gap can be calculated in a straightforward manner by inserting the field operators (\ref{eq:field_op}) into Eq. (\ref{eq:gap_def}) and then using the Bogoliubov transformation (\ref{eq:Bogolobov_trans}).

\section*{Appendix B: Elementary estimate of the superconducting nanowire parameters and additional physical comments}
A brief note concerning the experimental situation is in place here. The experimental evidence for superconductivity appearance in the thinnest nanowires of radius $R\sim1\:$nm in not clearly established\cite{Zgirski2005} even though for thicker nanowires the $T_C$ value grows with the decreasing diameter. In particular, the role of the order-parameter quantum fluctuations has been discussed as a detrimental factor\cite{Zgirski2008} in the case of zero field. Here we have considered relatively thin wires and have shown that a stable BCS or FF states can appear even for $R\sim1$nm and in strong fields. Such state of affairs can be understood within the following semi-macroscopic picture emerging from the Ginzburg-London type of approach\cite{Cyrot1992} in which we treat the nanowire as a 3-dimensional cylinder filled with electron gas and do not include the quantum size effect explicitly. The London penetration depth $\lambda_L=(m_0c^2/e^2n_c)^{1/2}$ is for the low carrier concentration $n_c\sim4\cdot10^{18}\:$cm$^{-3}$ of 
the order of $10^3\:$nm, i.e., much larger than the wire diameter. On the other hand, the coherence length $\xi=\hbar^2k_F/(m^*\Delta)$, where 
$k_F$ is the Fermi 
wave vector, is much smaller than $\lambda_L$ and is of the order of $R$. Hence, we have the superconductor of extreme type II. Note that bulk Al is the superconductor of type I. The difference appears because for the wire the carrier concentration is about 4 orders of magnitude smaller and in this situation, the values of the $\lambda_L\sim 1/n_c$ and $\xi\sim\hbar k_F\sim n_c^{1/3}$ will be vastly different. It should also be noted that in nanostructures the coherence length can be suppressed even more due to the quantum size effect\cite{Shanenko2010}. Under the circumstance that the field penetrates the whole sample, we can estimate the second critical field, which is $H_{c2}=\Phi_0/2\pi\xi^2$, where $\Phi_0$ is the flux quantum. Explicitly, $H_{c2}\sim10^2\:$T, whereas the first critical field $H_{c1}\equiv\Phi_0/(4\pi\lambda^2_L)\ln(\xi/\lambda_L)\sim 10^{-3}$T. Note that the fields, at which the FF state appears (cf. Figs 3(a), 5, and 6) are in the similar interval, explicitly $10-100\:$T. 
Therefore, one may ask whether the flux lines do not form along the wire direction instead of the FF state. This seems not feasible because the wire is long and 
too thin as we have assumed at the start. Explicitly, the size of the London orbit is $l_0=(\Phi_0/2\pi H_a)^{1/2}$ and equals to $l_0=3\:$nm and $2\:$nm for $H_a=50\:$T and $100\:$T, respectively. In such case the FF state is formed, with the supercurrent density $j_s=|e|n_c(\hbar q/m*)$ and stabilizes the superconducting state, as the Cooper pair total momentum compensates the Fermi wave vector mismatch. Taking $q\sim0.01\:$nm$^{-1}$ (cf. Fig. \ref{fig3}c) we obtain the supercurrent $I=j_s \pi R^2$, which for $R=1\:$nm amounts to the value $\sim 1\:\mu$A. The pair velocity in FF state is $\hbar q/m^*\sim10^3\:$m/s, whereas the Fermi velocity is about $7\cdot 10^5\:$m/s, so the velocity of a steady Cooper-pair flow is much smaller than the Fermi velocity. One should note that such current in turn produces the magnetic field at the wire surface of the magnitude $B=\mu_0I/(2\pi R)\sim10^{-4}\:$T, i.e., far below the order of magnitude of the applied field. As a result, the magnetic field induced by the supercurent in the FF 
state does not compensate the applied field which causes the Fermi wave vector mismatch. Obviously, this is a mean-field (BCS-like) picture which should most probably be 
amended by the discussion of the order-parameter fluctuation effects, since the coherence length is rather small. Nonetheless, the present discussion provides a coherent overall picture, as well as justifies our assumption that the field is uniform within the sample volume. Obviously, we assume that the wire cross section is uniform, presumably monocrystalline, and in the clean limit. Perhaps, the ring configuration would be the best for experiments, since in such geometry the Fulde-Ferrell phase is more appropriate than a formation of the Larkin-Ovchinnikov supercurrent standing-wave state.

Parenthetically, the physical situation is reminiscent of what 
takes place in the case of high-temperature superconductors. It is thus tempting to suggest an experiment on wires (or rather, thin stripes) composed of a high-temperature superconductor.


%

\end{document}